\documentclass[prb,reprint,twocolumn,superscriptaddress,noshowpacs,notitlepage,longbibliography,10pt,citeautoscript,floatfix]{revtex4-2}%
\usepackage{graphicx,bm,times}
\usepackage[dvipsnames]{xcolor}
\usepackage{booktabs}
\usepackage{lipsum}
\usepackage{amsmath}
\usepackage{amsfonts}
\usepackage{amssymb}
\usepackage{mathtools}
\usepackage{color}
\usepackage{hyperref}
\usepackage{chemformula}
\usepackage{acro}
\hypersetup{colorlinks=true,allcolors={blue}}
\DeclareAcronym{DFT}{short = DFT, long = density functional theory}
\DeclareAcronym{REXS}{short = REXS, long = resonant elastic X-ray scattering}
\DeclareAcronym{AFQ}{short = AFQ, long = antiferroquadrupolar}
\DeclareAcronym{FQ}{short = FQ, long = ferroquadrupolar}
\DeclareAcronym{AFM}{short = AFM, long = antiferromagnetic}
\DeclareAcronym{ATS}{short = ATS, long = anisotropic tensor susceptibility}
\DeclareAcronym{BZ}{short = BZ, long = Brillouin zone}
\DeclareAcronym{PDOS}{short = PDOS, long = projected density of states}
\DeclareAcronym{DOS}{short = DOS, long = density of states}
\DeclareAcronym{irrep}{short = irrep, long = irreducible representation}
\begin{document}
\title{Antiferroquadrupolar Order in Altermagnetic \ch{CoF2}}
\author{Daniel Halliday}
\email{daniel.r.halliday@durham.ac.uk}
\affiliation{Department of Physics, Durham University, South Road, Durham DH1 3LE, United Kingdom}
\author{Laura P\"oysti}
\affiliation{Department of Physics and Astronomy, University College London, Gower Street, London WC1E 6BT, United Kingdom}
\author{Chung Xu}
\affiliation{Department of Physics and Astronomy, University College London, Gower Street, London WC1E 6BT, United Kingdom}
\author{Daniel A. Mayoh}
\affiliation{Department of Physics, University of Warwick, Gibbet Hill Road, Coventry, CV4 7AL, United Kingdom}
\author{Didier Wermeille}
\affiliation{XMaS - UK CRG Beamline, European Synchrotron Radiation Facility (ESRF), 71 avenue des Martyrs, 38000 Grenoble, France}
\author{Dharmalingam Prabhakaran}
\affiliation{Department of Physics, University of Oxford, Parks Road, Oxford, OX1 3PU, United Kingdom}
\author{David R. Bowler}
\affiliation{Department of Physics and Astronomy, University College London, Gower Street, London WC1E 6BT, United Kingdom}
\affiliation{London Centre for Nanotechnology, 17-19 Gordon St, London WC1H 0AH, United Kingdom}
\author{Roger D. Johnson}
\email{roger.d.johnson@durham.ac.uk}
\affiliation{Department of Physics, Durham University, South Road, Durham DH1 3LE, United Kingdom}
\date{\today}
\begin{abstract}
Altermagnets host non-relativistic spin-split electronic states whose microscopic origin is theoretically linked to a hidden charge order, yet experimental studies of this charge order are limited. We therefore investigate charge order within \ch{CoF2}, a $d$-wave altermagnetic compound with a rutile crystal structure and $\Gamma$-point antiferromagnetism. Combining resonant elastic X-ray scattering, symmetry analysis and \emph{ab initio} calculations, we directly observe charge ordering and identify it as antiferroquadrupolar in nature. Via electronic structure calculations, we show that the experimentally observed antiferroquadrupolar order gives rise to the characteristic altermagnetic spin-splitting, thereby establishing empirical evidence for the decomposition of the altermagnetic order parameter into magnetic and charge degrees of freedom. We hence demonstrate that antiferroquadrupolar order is the microscopic origin of altermagnetism in \ch{CoF2}, with implications to the wider family of rutile altermagnets. Furthermore, our approach is applicable to studying altermagnetism in general, having demonstrated that resonant elastic X-ray scattering can serve as a direct probe of the charge multipoles that underpin spin-split electronic states in these materials.
\end{abstract}
\maketitle
Altermagnets are a recently classified group of magnetic materials, which have no net magnetic moment but display time reversal symmetry ($\mathcal{T}$) breaking phenomena \cite{Smejkal2022}. Such $\mathcal{T}$ breaking phenomena have been experimentally observed in a small number of materials, for instance, a spin-splitting of the electronic states \cite{Roig2024,Krempasky2024,Lee2024}, the anomalous Hall effect \cite{Reichlova2020,Betancourt2023,Reichlova2024} and the splitting of magnons into modes of opposite chirality \cite{Liu2024,Sun2025,Singh2026}. Excitingly, altermagnetic materials hold significant potential for applications in spintronic devices \cite{Smejkal2022a} due to exotic phenomena they display, including spin-current generation \cite{Naka2021,Sourounis2025} and ultra-fast magneto-resistive switching \cite{Smejkal2022b}. Consequently, a great scientific effort has been directed into the exploration of potential altermagnetic candidates. However, despite the large number of predicted altermagnets, the number of materials with experimentally confirmed altermagnetic signatures remains low. One family of altermagnetic candidate materials that have been the subject of particularly intense study are those with a rutile crystal structure (chemical formula $TX_2$, where $T$ is a divalent or tetravalent transition metal cation, and $X$ is an anion, respectively) \cite{Adamantopoulos2024,Samanta2025,Faure2025,Yuan2020,Sears2026}. Initially, the rutile compound \ch{RuO2} was identified as a promising altermagnetic candidate \cite{Ahn2019,Feng2022,Fedchenko2024,Smejkal2020}, however, recent studies have reported the absence of long-range magnetic order in this material \cite{Hiraishi2024,Kesler2024,Song2025,Occhialini2026,Wang2026}, thereby excluding it as an altermagnet. The investigations surrounding \ch{RuO2} have highlighted the need for well-defined criteria for characterising the order parameter of altermagnets. 

Theoretical studies have proposed that the altermagnetic order parameter may be described in terms of a ferroic ordering of local multipolar moments that are macroscopically $\mathcal{T}$-odd~\cite{Hayami2019,Winkler2023,Radaelli2024}. For rutile altermagnets of so-called $d$-wave symmetry, this is an ordering of magnetic octupoles with order parameter, $\mathcal{O}$ \cite{Bhowal2024}, which can be constructed through the direct product of a magnetic dipole order parameter, $\mathcal{L}$, and a charge quadrupolar order parameter, $\mathcal{Q}$ \cite{McClarty2024}
\begin{equation}\label{altermagnetic order parameter}
    \mathcal{O} = \mathcal{L} \otimes \mathcal{Q}.
\end{equation}

Generally, $\mathcal{L}$ is only present for $T < T_N$ and $\mathcal{Q}$ is persistent at all temperatures. Presently, experimental studies investigating this multipolar ordering of altermagnets, in addition to the decomposition into magnetic and charge components, are somewhat lacking \cite{Gyo2025,Usachev2026}. Therefore, in this letter, we present a combined experimental and theoretical study of the multipolar order of the rutile altermagnet \ch{CoF2}. \Ac{REXS} measurements reveal the charge quadrupole order ($\mathcal{Q}$), with its symmetry determined via theoretical modeling of our data. We then present \ac{DFT} calculations, which demonstrate that, when combined with the collinear \ac{AFM} order ($\mathcal{L}$), it produces an altermagnetic spin-splitting of the electronic states consistent with the octupolar order parameter ($\mathcal{O}$) predicted to occur in rutile-type altermagnets.

\begin{figure}[t]
  \centering
  \includegraphics[width=\columnwidth]{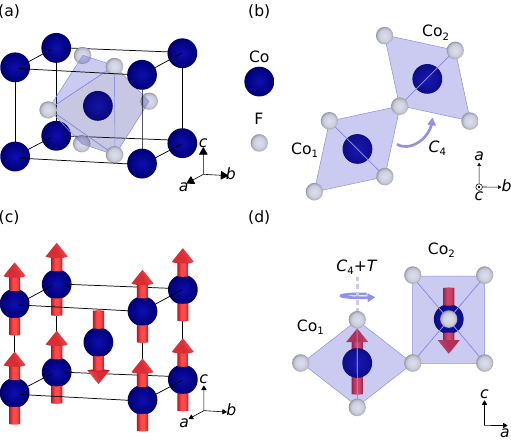}
  \caption{(a) A single unit cell of \ch{CoF2} with lattice vectors labeled. (b) The two \ch{Co} sites within the primitive unit cell. \ch{Co_1} sits at [0,0,0] and \ch{Co_2} at [$\frac{1}{2}$,$\frac{1}{2}$,$\frac{1}{2}$] \cite{OToole2001}. These two \ch{Co} sites are related by a $4_2$ screw; a rotation around the $c$-axis by 90$^\circ$ followed by a translation along the $c$-axis by $\frac{c}{2}$. (c) \ac{AFM} ground state of \ch{CoF2}, the direction of the magnetic moments is shown by the red arrows. \ch{F} atoms have been removed for clarity (d) The two opposite spin sub-lattices of \ch{CoF2}, \ch{Co_1} and \ch{Co_2}. A combination of the $4_2$ screw and time-reversal symmetry, $\mathcal{T}$, transforms from one sub-lattice to the other.}
  \label{fig:figure_1_main}
\end{figure}

Shown in figure \ref{fig:figure_1_main}a is the crystal structure of \ch{CoF2} (space group $P4_2/mnm$ \cite{OToole2001}), with two \ch{Co} ions located at the origin and body-center, coordinated by a \ch{F6} octahedron. The two \ch{CoF6} octahedra are related by a $4_2$ screw parallel to the $c$-axis (as illustrated in figure \ref{fig:figure_1_main}b) and an $n$-glide operation. On cooling below $T_\mathrm{N} \approx 37$~K, \ch{CoF2} develops long-range \ac{AFM} order \cite{Renzi1984,Chatterji2009}. The magnetic moments of the two Co ions anti-align along the $c$-axis \cite{Chatterji2010}, as depicted in figure \ref{fig:figure_1_main}c, yielding a ground state symmetry described by the magnetic space group $P4_2^\prime /mnm^\prime$. The respective magnetic point group is $4^\prime /mm^\prime m$ (figure \ref{fig:figure_1_main}d), which satisfies the criteria for Type II altermagnetism \cite{Cheong2024} with a $d$-wave order parameter \cite{Guo2023}.

A single crystal of \ch{CoF2} was grown by the Bridgman technique \cite{Disa2020} and prepared for \ac{REXS} measurements by cutting a $(1,0,0)$ orientated face polished to a mirror finish. \ac{REXS} measurements along the $(h,0,0)$ zone axis were performed using the BM28 beamline (XMaS) at the European Synchrotron Radiation Facility \cite{Brown2001}. The \ch{CoF2} sample was measured at $T \sim 298$ K using incident X-rays tuned close to the \ch{Co} $K$-edge (7.709~keV) and linearly polarized perpendicular to the vertical scattering plane (labeled $\sigma$). Higher-order harmonics were rejected by a low-angle (4 mrad) reflection from \ch{Cr}-coated mirrors. The polarization of the scattered X-ray beam was analyzed using a pyrolytic graphite crystal scattering at the (0,0,6) Bragg reflection close to Brewster’s angle.

\begin{figure}
  \centering
  \includegraphics[width=\columnwidth]{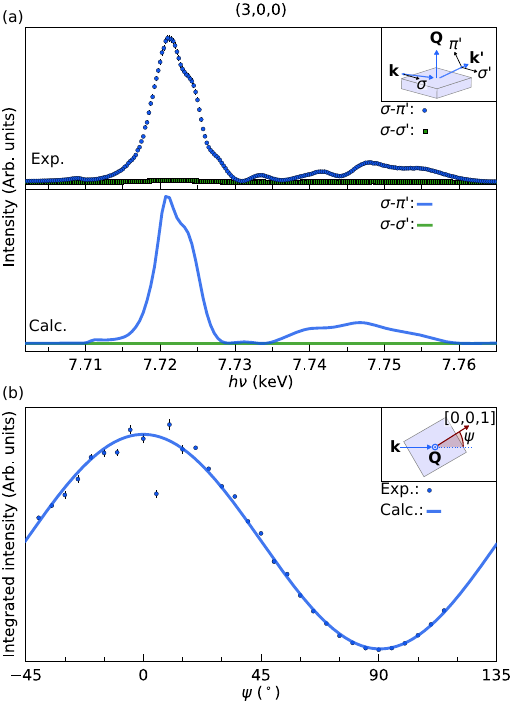}
  \caption{(a) Experimental (Exp.) and calculated (Calc.) photon-energy dependence of the \textbf{Q} $=\left(3,0,0\right)$ reflection within the: $\sigma$-$\pi^\prime$ channel (blue circles/line) and $\sigma$-$\sigma^\prime$ channel (green squares/line). The inset shows the scattering geometry for the \ac{REXS} experiment, where \textbf{k} is the incident wavevector, \textbf{k$^\prime$} is the scattered wavevector, \textbf{Q}$= \textbf{k}-\textbf{k$^\prime$}$ is the scattering vector and $\sigma$, $\sigma^\prime$/$\pi^\prime$ label the polarization of the incident and scattered X-rays, respectively. (b) Integrated intensity of the \textbf{Q} $=\left(3,0,0\right)$ reflection ($\sigma$-$\pi^\prime$ channel and $h\nu = 7.722$ keV) as a function of azimuthal angle $\left(\psi\right)$. As defined within the inset, $\psi$ is the angle between the [0,0,1] axis of \ch{CoF2} and the straight through X-ray beam when projected into the plane perpendicular to \textbf{Q}. The experimental data (Exp., blue circles) are described by a $\cos^2\psi$ function (Calc., blue line).}
  \label{fig:figure_2_main}
\end{figure}

At the reflection \textbf{Q} $= \left(3,0,0\right)$ of \ch{CoF2} we observe a sharp peak of scattered intensity (see supplementary figure 1a) and shown in figure \ref{fig:figure_2_main}a is a photon-energy scan of this reflection. Within \ch{CoF2}, the $\left(3,0,0\right)$ reflection is forbidden by symmetry, and in the absence of long-range magnetic order ($T > T_N$) must originate in so-called Templeton-Templeton \cite{Templeton1982} or \ac{ATS} \cite{Dmitrienko1983} scattering due to contrasting local charge distributions of the two \ch{Co} sites. The resonance is expected to originate in electric dipole transitions (E1-E1) that dominate the scattering cross-section, and hence corresponds to a transition from occupied \ch{Co} 1$s$ orbitals to unoccupied 4$p$ orbitals. We find that the main absorption edge at $\sim$7.72~keV appears approximately 10~eV above the tabulated \ch{Co} $K$-edge binding energy (supported by fluorescence measurements shown in supplementary figure 1b), likely due to interactions with nearest-neighbor \ch{F} orbitals. In addition to the main absorption edge, we find a small pre-edge feature and higher-energy resonances extending up to 40~eV above the edge. Measuring the resonance for orthogonal polarizations of scattered X-rays (labeled $\sigma^\prime$ and $\pi^\prime$, depicted within the inset of figure \ref{fig:figure_2_main}a), we observe that the resonant signal occurs only in the rotated $\sigma-\pi^\prime$ channel. We note that the weak intensity found in the $\sigma-\sigma^\prime$ channel ($\sim 1 \%$ of $\sigma-\pi^\prime$) is due to polarization analyzer inefficiencies.

To model the photon-energy dependence of this resonant spectrum, we performed calculations using the FDMNES code \cite{Bunau2009,Guda2015} implementing a Perdew and Zunger exchange-correlation potential. Calculations utilized the finite difference method, with a cluster radius of 7 \r{A}, and used as input the reported bulk crystal structure from Ref. \cite{OToole2001}. Only dipole-dipole (E1-E1) events were considered within calculations. As shown in figure \ref{fig:figure_2_main}a, our calculations reproduce the key features of the experimental photon-energy scan. Finite scattered intensity is calculated to only occur within the rotated polarization channel, and the pre-edge, main resonance and higher-energy peak shapes are well reproduced. Our calculations therefore confirm the E1-E1 origin of the full resonant spectrum. 

To elucidate the symmetry of the \ac{ATS} scattering, we measured the $\left(3,0,0\right)$ reflection as a function of azimuth angle $\left(\psi\right)$. These measurements used a photon-energy of $h\nu~=~7.722$ keV and were performed within the $\sigma-\pi^\prime$ channel. As depicted within the inset of figure \ref{fig:figure_2_main}b, $\psi$ is the angle between the [0,0,1] axis of \ch{CoF2} and the straight through X-ray beam when projected into the plane perpendicular to \textbf{Q}. When $\psi=0^\circ$ we measure the scattered intensity to be maximal, with it decreasing until a minimal value at $\psi=90^\circ$. This azimuthal dependence can be fit by a $\cos^2\psi$ function, shown in figure \ref{fig:figure_2_main}b by the solid blue line.

\begin{figure}
  \centering
  \includegraphics[width=\columnwidth]{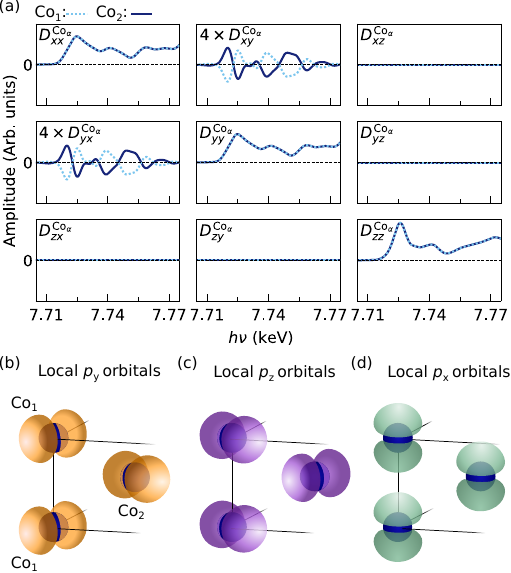}
  \caption{(a) Calculated photon-energy dependence of the matrix elements $D^{\ch{Co_{$\alpha$}}}_{ij}$ for: $\alpha =1$ (light blue dashed line) and $\alpha =2$  (dark blue line). Calculations are within the global coordinate system, $x\parallel\hat{a},y\parallel\hat{b},z\parallel\hat{c}$. Schematic of the: (b) and (c) antiferro-type ordering of the local $p_\mathsf{y}$ and $p_\mathsf{z}$ orbitals, respectively, and (d) the ferro-type ordering of the local $p_\mathsf{x}$ orbitals. The staggered orbital anisotropy of the local $p_\mathsf{y}$ and $p_\mathsf{z}$ orbitals corresponds to the \ac{AFQ} order present within \ch{CoF2}.}
  \label{fig:figure_3_main}
\end{figure}

Having measured the polarization and azimuthal dependence of the \textbf{Q} $= \left(3,0,0\right)$ reflection, we now consider the \ac{ATS} scattering tensor of \ch{CoF2} to determine the origin of this \ac{REXS} signal \footnote{Alternatively to the \ac{ATS} approach, one may utilize the tensorial approach of Lovesey \emph{et al.} \cite{Lovesey2005}, as outlined in the supplemental material \cite{SM}}. Within the global basis of the crystal $\left(x\parallel\hat{a},y\parallel\hat{b},z\parallel\hat{c}\right)$ we write the resonant tensor structure factor as
\begin{eqnarray}
    F_{\left(h,0,0\right)} &=& \sum_{ij} D^{\ch{Co_1}}_{ij} + D^{\ch{Co_2}}_{ij}, \quad h~\mathrm{even} \label{Resonant scattering tensor even}\\
    F_{\left(h,0,0\right)} &=& \sum_{ij} D^{\ch{Co_1}}_{ij} - D^{\ch{Co_2}}_{ij}, \quad h~\mathrm{odd} \label{Resonant scattering tensor odd}
\end{eqnarray}
where $D^{\ch{Co_{$\alpha$}}}_{ij}$ are the photon-energy dependent matrix elements of the E1-E1 scattering matrix for the two \ch{Co} sites ($\alpha = 1$ or 2). The good agreement between our calculations and experiment in figure \ref{fig:figure_2_main}a allows us to utilize FDMNES to evaluate these site-resolved matrix elements, as shown in figure \ref{fig:figure_3_main}a. We observe that the diagonal matrix elements are finite throughout the resonant energy range, and of equal amplitude for both \ch{Co_1} and \ch{Co_2}. On the contrary, the $ij = xy$ and $yx$ elements are finite but reverse sign between the different \ch{Co} sites, a consequence of the $4_2$ screw axis present within \ch{CoF2}. All other matrix elements are exactly zero by symmetry. Therefore, the diagonal elements enter into the structure factor for even $h$ (equation \ref{Resonant scattering tensor even}), while only the $ij = xy$ and $yx$ matrix elements contribute to the structure factor for odd $h$ (equation~\ref{Resonant scattering tensor odd}). Focusing on the $\left(3,0,0\right)$ reflection, equation \ref{Resonant scattering tensor odd} becomes
\begin{equation}\label{Resonant scattering tensor matrix}
F_{(3,0,0)} =
\begin{pmatrix}
\cdot & 2\chi & \cdot \\
2\chi & \cdot & \cdot \\
\cdot & \cdot & \cdot
\end{pmatrix}
\end{equation}
where we have used $\chi~=~D^{\ch{Co_1}}_{xy}~=~D^{\ch{Co_1}}_{yx}~=~-D^{\ch{Co_2}}_{xy}~=~-D^{\ch{Co_2}}_{yx}$. Given this form of the total resonant scattering tensor, we find that \cite{SM}
\begin{equation}\label{Resonant scattering intensity}
\begin{split}
I_{\left(3,0,0\right)}^{\sigma\pi^\prime}& \propto 4\chi^2\cos^2\theta_\mathrm{B} \cos^2\psi ,\\
I_{\left(3,0,0\right)}^{\sigma\sigma^\prime}& = 0
\end{split}
\end{equation}
where $\theta_\mathrm{B}$ is the Bragg angle of the reflection.

Equation \ref{Resonant scattering intensity} reproduces both the polarization and azimuthal dependencies we measured using \ac{REXS} (figure \ref{fig:figure_2_main}), and hence our analysis shows that resonant scattering at the $\left(3,0,0\right)$ reflection originates from a scattering tensor of pure $Q_{xy}$ symmetry (equation \ref{Resonant scattering tensor matrix}), where $Q_{xy}$ denotes a rank-2 electric quadrupole component. Moreover, the sign reversal of the tensor matrix elements between \ch{Co_1} and \ch{Co_2} sites (figure \ref{fig:figure_3_main}a) indicates the $Q_{xy}$ quadrupoles of the two sites have a staggered arrangement, known as \ac{AFQ} order.

\begin{figure}
  \centering
  \includegraphics[width=\columnwidth]{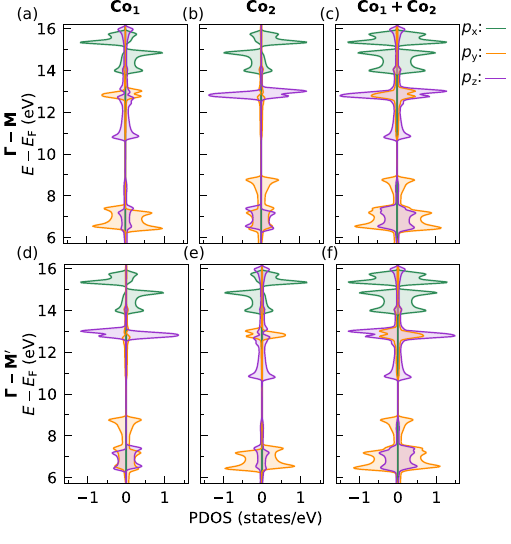}
  \caption{Site and spin resolved \ac{PDOS} of \ch{Co} for the local $p_\mathsf{x}$ (green), $p_\mathsf{y}$ (yellow) and $p_\mathsf{z}$ (purple) orbitals of: (a) \ch{Co_1} and (b) \ch{Co_2}, projected along the $\Gamma-\mathrm{M}$ direction. (c) Spin-polarized \acs{PDOS} of the local $p$ orbitals summed over both \ch{Co} sites, projected along the $\Gamma-\mathrm{M}$ direction. (d) to (f) are as (a) to (c), but projected along the orthogonal $\Gamma-\mathrm{M}^\prime$ direction.}
  \label{fig:figure_4_main}
\end{figure}

As the measured resonant process involves a transition from the occupied \ch{Co} $1s$ states into the unoccupied $4p$ states, it is natural to consider this \ac{AFQ} order within a local $p$ orbital basis of the \ch{Co} atoms. We note that while one might anticipate requiring $d$-orbitals for quadrupolar order, the charge density associated with a given $p$-orbital in fact carries a full quadrupolar moment \cite{SM}. We define a local $p$ orbital basis as
\begin{equation}\label{Local p orbitals Co1}
p^{\ch{Co_1}}_\mathsf{x} \parallel \hat{c}\text{, } p^{\ch{Co_1}}_\mathsf{y} \parallel \left(\hat{a}-\hat{b}\right)\text{, } p^{\ch{Co_1}}_\mathsf{z} \parallel \left(\hat{a}+\hat{b}\right)
\end{equation}
for \ch{Co_1} atoms, and
\begin{equation}\label{Local p orbitals Co2}
p^{\ch{Co_2}}_\mathsf{x} \parallel \hat{c}\text{,   } p^{\ch{Co_2}}_\mathsf{y} \parallel \left(\hat{a}+\hat{b}\right)\text{, } p^{\ch{Co_2}}_\mathsf{z} \parallel \left(-\hat{a}+\hat{b}\right)
\end{equation}
for \ch{Co_2} atoms, where the local coordinates are related by a $90^\circ$ rotation around the $c$-axis, consistent with the crystal symmetry of \ch{CoF2}. The $Q_{xy}$ quadrupole has lobes along the $\left(\hat{a}+\hat{b}\right)$ and $\left(\hat{a}-\hat{b}\right)$ directions, corresponding to the local $p_\mathsf{z}$, $p_\mathsf{y}$ orbitals of \ch{Co_1}, and the local $p_\mathsf{y}$, $p_\mathsf{z}$ orbitals of \ch{Co_2}, respectively. Therefore, the \ac{AFQ} order within \ch{CoF2} may be described as a staggered arrangement of the local $p_\mathsf{y}$ and $p_\mathsf{z}$ orbitals across the two \ch{Co} sub-lattices, as illustrated in figure \ref{fig:figure_3_main}b and c. Whereas the local $p_\mathsf{y}$ and $p_\mathsf{z}$ orbitals display an antiferro-type ordering, it follows from equations \ref{Local p orbitals Co1} and \ref{Local p orbitals Co2} that the local $p_\mathsf{x}$ orbitals of the two \ch{Co} sites have a ferro-type ordering, as illustrated in figure \ref{fig:figure_3_main}d. We note that these $p_\mathsf{x}$ orbitals give rise to \ac{ATS} scattering for $(h,0,0)$, $h$ even (equation \ref{Resonant scattering tensor even}). 

Having established a hidden quadrupolar order in \ch{CoF2}, we now examine its relation to altermagnetism. To do so, we performed electronic-structure calculations using \ac{DFT} implemented in the CONQUEST code (version 1.5) \cite{Bowler2002,Miyazaki2004,Nakata2022}. Calculations were performed using the Perdew–Burke–Ernzerhof exchange–correlation functional \cite{Perdew1996}, a Hubbard correction of $U = 2.25$~eV, and a triple zeta double polarized basis set \cite{Bowler2019,Hamann2013} generated using ONCVPSP-type pseudopotentials from the PseudoDojo library \cite{Hamann2013,VANSETTEN201839}. Structural relaxations were performed using a $4\times4\times6$ \textbf{k}-point mesh and a grid cut-off of $200$~Ha, resulting in relaxed lattice parameters within $\sim 1\%$ of those experimentally reported in Ref. \cite{OToole2001}. As our \ac{REXS} measurements have probed the unoccupied \ch{Co} $4p$ states, we focus our \ac{PDOS} calculations on these orbitals, utilizing the local orbital basis of the \ch{Co} sites introduced in equations \ref{Local p orbitals Co1} and \ref{Local p orbitals Co2}. In figure ~\ref{fig:figure_4_main}a and b, we show the calculated \ac{AFM} \ac{PDOS} for the two \ch{Co} sites evaluated along the $\Gamma$-M \ac{BZ} direction. For each site, we observe $\mathcal{T}$ breaking in all $p$ states, as expected. For a simple antiferromagnet, spin degeneracy would be recovered by summing over both \ch{Co} sites. However, performing this sum, as shown in figure~\ref{fig:figure_4_main}c, we find this degeneracy is not recovered for the local $p_\mathsf{y}$ and $p_\mathsf{z}$ states, indicating significant altermagnetic spin-splitting in these states. Conversely, in the sum, the $p_\mathsf{x}$ states are spin degenerate to good approximation. We note that in our DFT results there exists a small ($\sim1\%$) altermagnetic spin splitting of the local $p_\mathsf{x}$ states, which cannot be resolved in figure \ref{fig:figure_4_main}. This splitting is almost two orders of magnitude smaller than the largest spin splitting of $p_\mathsf{y}$ and $p_\mathsf{z}$. We argue that this residual splitting to some hybridization of the $p_\mathsf{x}$ states in our \ac{DFT} calculations, not considered in our minimal model introduce in figure \ref{fig:figure_3_main}d. Resolving the \ac{PDOS} along the orthogonal $\Gamma-\mathrm{M}^\prime$ \ac{BZ} direction (figure ~\ref{fig:figure_4_main}d and e), we again observe $\mathcal{T}$-breaking within the $p$ states. However, the sum over both \ch{Co} sites reveals that the spin-splitting of the local $p_\mathsf{y}$ and $p_\mathsf{z}$ is now inverted, figure ~\ref{fig:figure_4_main}f, exactly as required by $d$-wave altermagnetism \cite{Smejkal2022,Smejkal2022a}. Therefore, our \ac{DFT} calculations directly link our experimental \ac{REXS} results with altermagnetism present in the band structure. The \ac{AFQ} order of the $p_\mathsf{y}$ and $p_\mathsf{z}$ states, dressed by the \ac{AFM} order, gives rise to a non-relativistic spin-splitting which changes sign on rotating the crystal momentum by $90^\circ$. Unlike the antiferroically ordered $p_\mathsf{y}$/$p_\mathsf{z}$ states, the ferroically ordered $p_\mathsf{x}$ states remain largely spin-degenerate under the \ac{AFM} order, therefore we conclude that the ferro-type order does not give rise to altermagnetism. 

Finally, we employ symmetry arguments to relate our results back to the anticipated octupolar altermagnetic order parameter of \ch{CoF2}. The \ac{AFM} order transforms by the $m\Gamma_2^+$ \ac{irrep} of the parent paramagnetic group, while the \ac{AFQ} order transforms by the $\Gamma_1^+$ \ac{irrep} (totally symmetric under all translations and rotations of the crystal structure). The tensor product of these two yields $m\Gamma_2^+ \otimes \Gamma_1^+ = m\Gamma_2^+$. The transformation properties of $m\Gamma_2^+$ are exactly those required by the octupolar altermagnetic order parameter of \ch{CoF2} ($\mathcal{O}$), and hence the \ac{AFQ} order should be understood as a manifestation of the hidden charge order ($\mathcal{Q}$) that results in $d$-wave altermagnetism in rutile compounds.

In summary, via resonant X-ray scattering measurements and \emph{ab initio} calculations, we have characterized an underlying \ac{AFQ} order within \ch{CoF2}, and demonstrated that this order is central to generating $d$-wave altermagnetic spin-splitting. Our investigation provides crucial experimental evidence for the decomposition of the altermagnetic order parameter of \ch{CoF2} into charge and magnetic degrees of freedom. Our results may be applied directly to the wider family of altermagnetic materials which share the rutile crystal structure and $\Gamma$-point \ac{AFM} order \cite{Lovesey2026}, and more broadly, demonstrate \ac{REXS} as an ideally suited experimental probe of multipolar order in all altermagnets.\par
\emph{Acknowledgments $-$}
We acknowledge funding through EPSRC Grant UKRI648. DH thanks Prof. T. Hase and Dr. Y. Joly for useful discussions. LP thanks Prof. T. Miyazaki for useful discussions. We acknowledge the European Synchrotron Radiation Facility (ESRF) for provision of synchrotron radiation facilities under proposal ID A28-1-1496 and on beamline BM28 (XMaS) \cite{XMaS}. Part of this work has made use of the Hamilton HPC Service of Durham University and the HPC Clusters of the London Centre for Nanotechnology.
\bibliography{ref.bib}
\end{document}